\documentclass[pra,aps, twocolumn,groupedaddress]{revtex4}
\usepackage{amssymb,amsmath,amsfonts,epsfig,graphicx,latexsym,bm,color}

\newcommand{\beq}{\begin{eqnarray}}
\newcommand{\eeq}{\end{eqnarray}}
\newcommand{\bem}{\begin{pmatrix}}
\newcommand{\eem}{\end{pmatrix}}
\newcommand{\nn}{\nonumber}
\newcommand{\hb}{\hbar}

\newcommand{\f}{\frac}
\newcommand{\rfs}[1]{Eq.~(\ref{#1})}

\newcommand{\h}[1]{\hat{#1}}
\newcommand{\dr}[1]{|#1\rangle}

\newcommand{\tb}[1]{\textbf{#1}}
\newcommand{\tr}[1]{\textrm{#1}}
\newcommand{\ro}[1]{\sqrt{#1}}
\newcommand{\ex}[1]{\langle #1 \rangle}
\def\nn{\nonumber}

\def\e{\epsilon}
\def\d{\delta}

\def\l{\lambda}
\def\o{\omega}
\def\O{\Omega}
\def\p{\phi}
\def\B{\beta}

\def\A{\alpha}
\def\t{\tau}
\def\s{\sigma}
\def\th{\theta}
\def\lt{\left}
\def\rt{\right}

\begin{document}
\title{Artificial Staggered Magnetic Field for Ultracold Atoms in Optical Lattices}
\author{Lih-King Lim$^{1}$, Andreas Hemmerich$^{2}$, and C. Morais Smith$^{1}$}
\affiliation{$^{1}$Institute for Theoretical Physics, Utrecht University,
Leuvenlaan 4, 3584 CE Utrecht, The Netherlands}
\affiliation{$^{2}$Institut f\"{u}r Laser-Physik, Universit\"{a}t Hamburg,
Luruper Chaussee 149, 22761 Hamburg, Germany}
\date{\today}

\begin{abstract}
A time-dependent optical lattice with staggered particle current in the tight-binding regime was considered that can be described by a time-independent effective lattice model with an artificial staggered magnetic field. The low energy description of a single-component fermion in this lattice at half-filling is provided by two copies of ideal two-dimensional massless Dirac fermions. The Dirac cones are generally anisotropic and can be tuned by the external staggered flux $\p$. For bosons, the staggered flux modifies the single-particle spectrum such that in the weak coupling limit, depending on the flux $\p$, distinct superfluid phases are realized. Their properties are discussed, the nature of the phase transitions between them is establised, and Bogoliubov theory is used to determine their excitation spectra. Then the generalized superfluid-Mott-insulator transition is studied in the presence of the staggered flux and the complete phase diagram is established. Finally, the momentum distribution of the distinct superfluid phases is obtained, which provides a clear experimental signature of each phase in ballistic expansion experiments.
\end{abstract}

\maketitle
\date{\today}

\section{Introduction}
The preparation of clean condensed-matter systems is typically limited by disorder resulting from inevitable impurities, and relevant
physical parameters often cannot be controlled to high precision. In contrast, ultracold gases confined in optical lattices can be
controlled to perfection, which permits stringent confrontations between experiments and many-body theory. A prominent example is the superfluid-Mott-insulator transition in the Bose-Hubbard model in two \cite{Spielman:07} and three dimensions \cite{Greiner:02}, where experiments have provided a unique quantitative test ground for the respective theoretical predictions \cite{Fisher:89, Jaksch:98}. Recently, major efforts have been focused on reaching the quantum degenerate regime of the fermionic Hubbard model with ultracold atoms \cite{Schneider:08}, with the hope to promote our
present understanding of strongly correlated electronic systems (e.g., high-$\tr{T}_c$ superconductors).

The remarkable versatility of optical potentials should allow for the realization of the exotic physics known to occur for lattice electrons in strong magnetic fields. Until recently, the generation of artificial gauge fields for neutral atoms has been limited to spinning up the entire system, thereby mimicking the Lorentz force as experienced by a charge particle subjected to a magnetic field \cite{Matthews:99, Williams:99, Madison:00, Fetter:09}. For such systems, the regime of strong correlations has been shown to be very rich \cite{Cooper:01}. Reaching this regime, however, remains a technical challenge due to the requirement of rotation frequencies on the order of the trapping frequencies. Realizations of artificial gauge fields, which do not rely on large-scale rotations, have been proposed in a variety of theoretical works \cite{Juzeliunas:04, Lew:08, Gunter:09, Spielman:09}. The recent experimental demonstration of a light-induced artificial magnetic field by Lin \textit{et al.} \cite{Lin:09}, and most recently its application to excite a vortex lattice, has been a first step to overcome the limitations imposed by schemes based upon large scale rotation \cite{Lin:09b}.

In conventional solids, the creation of a magnetic flux strength on the order of a flux quantum $\Phi_0=h/e$ through a plaquette is a yet-unaccomplished challenge, which has impeded access to the rich physical regime characterized by the famous Hofstadter butterfly single-particle spectrum \cite{Hofstadter:76}.  Experiments have thus been limited to artificial superlattices with lattice constants on the order of 100 nm,
where in fact indications of a fractal energy spectrum could be observed \cite{Alb:01}. In optical lattices, it should be possible to achieve artificial magnetic fields with a magnetic length comparable to the lattice length scale. Nontrivial topological properties, such as the fractional quantum Hall effect \cite{Sorensen:05} and the anomalous quantum Hall effect \cite{Goldman:09b}, may thus become accessible. Theoretical studies of optical lattices with an artificial uniform magnetic field \cite{Jaksch:03} and their generalization to a non-Abelian gauge field \cite{Goldman:09a}, where phenomena such as the Escher ``staircase" \cite{Mueller:04} and the Hofstadter ``moth" \cite{Osterloh:05} are predicted, have attracted broad interest because experimental realizations with ultracold atoms may be possible.

Finally, because of their slow motional time scale, cold atoms in optical lattices are well suited for precise manipulation of the lattice dynamics by external driving \cite{Eckardt:05, Hem:07, Lim:08, Creffield:08, Zenesini:09}. The theoretical predictions of coherent control in an optical lattice with a time-periodic optical potential using Floquet theory \cite{Eckardt:05} were successfully tested in a recent experiment \cite{Zenesini:09}. These studies have shown that a temporal modulation, which acts to shake the lattice, can be used to modify the effective tunneling strength and even to tune it into the regime of negative values. Driven tunneling has also been studied for cold atoms subjected to double well potentials; phenomena predicted long ago, such as coherent destruction of tunneling \cite{Hae:91}, have been recently observed \cite{Obe:08}.

This paper discusses how driven tunneling can be used to generate an \textit{artificial staggered magnetic field} for atoms in a two-dimensional square optical lattice. A detailed description of the new physical phenomena that arise when the lattice is loaded with bosons is presented, thus extending a recent publication \cite{Lim:08}. In addition, a discussion for fermions is included. The paper is organized as follows. In Sec.~\ref{sec2}, two different methods are used to show how the time-dependent problem yields a time-independent effective lattice model with a staggered magnetic field. Section~\ref{sec3} describes a study of the effective Hamiltonian when the lattice is loaded with single-component fermionic atoms. The low-energy excitations at half-filling are shown to behave like Dirac particles. The anisotropic Dirac cones are discussed, the slope of which is tunable via the strength of the staggered flux. At $\pi$ flux we obtain the $\pi$-flux phase \cite{Affleck:88}. Section~\ref{sec4} contains a study of the generalized Bose-Hubbard model in the presence of staggered flux at zero temperature. In the weak coupling limit, depending on the flux $\p$, distinct superfluid phases are realized: a homogeneous zero-momentum superfluid; a staggered-vortex superfluid, characterized by a vortex-antivortex lattice with one vortex per plaquette; and a staggered-sign superfluid with an order parameter with opposite sign for adjacent lattice sites. The nature of the phase transitions between the different superfluid phases is established via a Hartree ansatz, and their excitation spectra are studied using Bogoliubov theory. In Sec.~\ref{sec5}, the superfluid-Mott-insulator transition in the strong coupling regime is determined in two different ways. The staggered flux renormalizes the phase boundary, and, thus, the generalized phase diagram is obtained with respect to the chemical potential, the onsite interaction, and the strength of the artificial magnetic field. In Sec.~\ref{sec6}, the distinct momentum distributions of the different superfluid phases are calculated, which allow for their discrimination in standard ballistic expansion experiments. Finally, the paper closes with conclusions in Sec.~\ref{conc}.

\section{Time-Dependent Optical Lattice and the generalized Hubbard model}\label{sec2}
\subsection{Time-dependent Hubbard model}
\begin{figure}
\includegraphics[scale=.35, angle=0, origin=c]{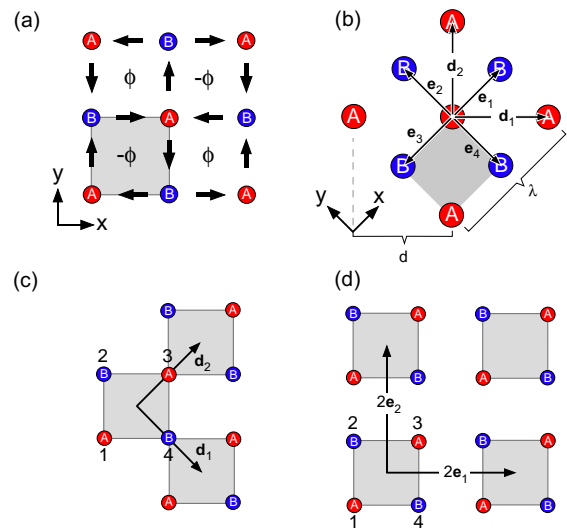}
\caption{\label{Fig.1} (Color online) (a) Schematic of the staggered current (black arrows) driven by the time-dependent optical lattice, which leads to two inequivalent sublattices $\mathcal{A}$ and $\mathcal{B}$. (b) $\tb{d}_1$ and $\tb{d}_2$ are the unit vectors of the $\mathcal{A}$ sublattice with length $d=\l/\sqrt{2}$, and $\tb{e}_l,\,\, l=1,2,3,4$ are the nearest-neighbor vectors connecting the sublattices. In (c) and (d), two distinct plaquette summation conventions are defined, denoted by $\sum_{\lozenge}$ and $\sum_{\square}$ in the text. In (c) the lattice is composed by translating an elementary plaquette (shaded area) by means of the primitive vectors $\tb{d}_1$ and $\tb{d}_2$ of the $\mathcal{A}$ sublattice. In (d) plaquettes are translated by the vectors $2\tb{e}_1$ and $2\tb{e}_2$. The four corners of an elementary plaquette are numbered consecutively by 1,2,3,4 as shown.}
\end{figure}

The starting point is the proposal of Ref.~\cite{Hem:07}, which pointed out that a refined modulation technique can be employed to induce an orbital current with a $d_{x^2-y^2}$ symmetry in a two-dimensional optical lattice. The optical potential takes the form $V(\tb{r},t)=V_0(\tb{r})+V_1(\tb{r},t)$, consisting of a stationary part $V_0(\tb{r})$ and a temporal modulation $V_1(\tb{r},t)$ with
\beq
V_0(\tb{r})&=&- \bar V_0 \rho(\tb{r}),\nn\\
V_1(\tb{r},t)&=& \kappa V_0(\tb{r})\cos(2 S(\tb{r})-\O t),
\eeq
where $\rho(\tb{r})=\sin^2(2\pi x/\l)+\sin^2(2\pi y/\l)$,
\beq
S(\tb{r})=\tan^{-1}\biggl\{\f{\sin(2\pi x/\l)-\sin(2\pi y/\l)}{\sin(2\pi x/\l)+\sin(2\pi y/\l)}\biggr\},
\eeq
$\l$ is the wavelength of the laser light, $\bar V_0$ is the mean well depth of the square lattice potential, $\Omega$ is the rotation frequency, and $\kappa$ is a parameter that quantifies the admixture of the temporal modulation term. It has been shown in Ref.~\cite{Hem:07} that this optical potential can be engineered in experiments by superimposing two bichromatic optical standing waves such that $\bar V_0$ can be varied between zero and hundreds of
the recoil energy $E_{\tr{R}} \equiv 2 \pi^2 \hb^2/m \l^2$, where $m$ denotes the mass of the atoms and $\kappa$ can be adjusted within the interval $[0,1]$. The stationary component $V_0(\tb{r})$ of the optical potential provides a regular square lattice potential with spacing $\l/2$, whereas the temporal modulation term $V_1(\tb{r},t)$ induces local rotation around each plaquette with opposite directions for neighboring plaquettes [see Fig.~\ref{Fig.1}(a)]; that is, it drives a staggered current that possesses $(d_{x^2-y^2})$-like symmetry.

Given the time-dependent confining potential $V(\tb{r},t)$, the Hamiltonian in the second quantized form describing the quantum gas in the ultracold regime can be written as
\beq\label{ham1}
H(t)&=&\int d^2 \tb{r} \,\psi^\dag (\tb{r})\lt(-\f{\hb^2}{2m}\nabla^2+V(\tb{r},t)\rt)\psi(\tb{r})\nn\\
&&+\f{1}{2}\f{4\pi a_s\hb^2}{m}\int d^2 \tb{r} \psi^\dag(\tb{r})\psi^\dag(\tb{r})\psi(\tb{r})\psi(\tb{r}),
\eeq
where $\psi(\tb{r})$ describes a bosonic (fermionic) field obeying commutation (anticommutation) relations, $m$ is the mass of the bosons (fermions), and $a_s$ is the $s$-wave scattering length. Following Ref.~\cite{Jaksch:98}, we write the atomic field operator $\psi(\tb{r})$ in terms of the Wannier wave functions $\psi(\tb{r})=\sum_i w(\tb{r}-\tb{R}_i)a_i$, where $\tb{R}_i$ denotes the potential minima of $V_0(\tb{r})$, at which the atoms are localized. The corresponding annihilation (creation) operator is denoted by $a_i \,(a^\dag_i)$. The staggered rotation yields a decomposition of the square lattice into two sublattices $\mathcal{A}$ and $\mathcal{B}$ [see Fig.~\ref{Fig.1}(b)]. The Bravais lattice is then given by one of the sublattices $\mathcal{A}$ or $\mathcal{B}$ and the unit cell is spanned by the lattice unit vectors [see Fig.~\ref{Fig.1}(b)]
\beq
\tb{d}_1= \tb{e}_1 +  \tb{e}_4,  \tr{\ \ \ }  \tb{d}_2= \tb{e}_1 +  \tb{e}_2,
\eeq
with the lattice constant $d=\l/\sqrt{2}$. The four vectors $\tb{e}_l \,\,( l=1,2,3,4)$, connecting an $\mathcal{A}$ site to its four nearest neighboring $\mathcal{B}$ sites, are defined by
\beq
\tb{e}_1=-\tb{e}_3 = \frac{\l}{2} \, \hat x, \tr{\ \ \ }\tb{e}_2=-\tb{e}_4 = \frac{\l}{2} \, \hat y,
\eeq
where $\hat x$ and $\hat y$ are the unit vectors in $x$ and $y$ directions shown in Fig.~\ref{Fig.1}. In order to distinguish the two sublattices in our description, we introduce two sets of annihilation (creation) operators, $a_i$ ($a_i^\dag$) and $b_i$ ($b_i^\dag$), corresponding to the operation on site $i$ of the sublattice $\mathcal{A}$ and $\mathcal{B}$, respectively. By substituting the Wannier expansion into \rfs{ham1}, we obtain the well-known Hubbard model with additional time-dependent one-body terms:
\beq\label{Ham2a}
H(t)&=& -\sum_{\tb{r}\in \mathcal{A}}\sum_{l=1}^4 J_0 (a_{\tb{r}}^\dag b_{\tb{r}+e_l}+\tr{H.c.})\nn\\&&+\f{1}{2}U_0\sum_{\tb{r}\in \mathcal{A} \oplus \mathcal{B}}n_{\tb{r}}\,\,(n_{\tb{r}}-1)\nn\\
&&+\chi_1 \sin (\O t)\sum_{\tb{r}\in \mathcal{A}}\sum_{l=1}^4(-1)^{l+1} (a^\dag_{\tb{r}}b_{\tb{r}+e_l}+ \tr{H.c.})\nn\\
&&+\chi_2 \cos(\O t) \sum_{\tb{r}\in \mathcal{A}} \biggl(n_{\tb{r}}- n_{\tb{r}+e_1}\biggr),
\eeq
where $n_{\tb{r}}$ is the number operator on site $\tb{r}$. The first two terms describe the well-known Hubbard model with the nearest-neighbor hopping energy $J_0$ and the onsite interaction strength $U_0$ given in terms of microscopic parameters in the standard manner, $J_0 = -\int d^2 \tb{r} \, w^*(x+\lambda/4,y+\lambda/4) [-(\hb^2/2m)\nabla^2+V_0(\tb{r})]w(x-\lambda/4,y+\lambda/4)$ and $U_0 = \sqrt{8\pi} (\hb^2 a_s/m \sigma_z) \int d^2\tb{r} \, |w(\tb{r})|^4$. Here, harmonic confinement of the atoms in the third direction is assumed with a localization radius $\sigma_z$. The modulation amplitudes are given by $\chi_1= \kappa \bar V_0 \int d^2 \tb{r} \, w^*(x+\lambda/4,y)[\sin^2(2\pi x/\l)-\cos^2(2\pi y/\l)]w(x-\lambda/4,y)$ and $\chi_2=2 \kappa \bar V_0 \int d^2 \tb{r} \, |w(\tb{r})|^2 \cos(2\pi x/\l)\cos(2\pi y/\l)$. Note that in Eq.~(\ref{Ham2a}) we have made the assumption that a single-band description, with all atoms residing in the lowest band, is sufficient. This requires the rotor frequency $\O$ to be detuned from interband resonance transitions of the system.

For the calculations that follow, it is convenient to rewrite the Hamiltonian (\ref{Ham2a}) as
\beq
\label{Ham2}
H(t) &=& H_0 + W(t) + H_{\tr{int}}, \nn \\ \nn \\
H_0 &=& -J_0 \mathcal{T}, \,\, \mathcal{T} \equiv \sum_{<i,j>}a_i^\dag b_j , \nn \\ \nn \\
W(t) &=& Q^{\dag} e^{i\O t} + Q e^{-i\O t}, \,\, Q \equiv \f{1}{2}\biggl(\chi_2\mathcal{N}+ i \chi_1\mathcal{M}\biggr), \nn \\  \nn \\
\mathcal{M}&\equiv&\sum_{\tb{r}\in \mathcal{A}, l=1-4}(-1)^{l+1} (a^\dag_{\tb{r}}b_{\tb{r}+e_l}+ \tr{H.c.}),  \nn\\
\mathcal{N}&\equiv& \sum_{\tb{r}\in \mathcal{A}} (n_{\tb{r}}-n_{\tb{r}+e_1}), \nn \\
H_{\tr{int}}&=& \f{1}{2}U_0\sum_{\tb{r}\in \mathcal{A} \oplus \mathcal{B}}n_{\tb{r}}\,(n_{\tb{r}}-1),
\eeq
where $H_0$ is the stationary kinetic term and $H_{\tr{int}}$ is the two-body onsite interaction. As is evident from the anisotropy of $W(t)$, the time-dependent part of the optical potential $V_1(\tb{r},t)$ renders the two sublattices inequivalent: an anisotropic (quadrupole-like) time modulation of the nearest-neighbor hopping ($\mathcal{M}$) and of the local chemical potential ($\mathcal{N}$) arises with a $\pi/2$ relative temporal phase lag, which introduces an alternating rotational sense to adjacent plaquettes.

\subsection{Effective staggered magnetic field}\label{sec2b}
In the following, we discuss two different approaches to obtain an effective time-independent description of the time-dependent Hamiltonian~(\ref{Ham2}). We begin with an expansion of the time-evolution operator of the one-body Hamiltonian $H_{\tr{1B}}(t) = H_0 + W(t)$ in a Dyson series. Making use of its temporal periodicity and neglecting higher-order many-body terms, we obtain an effective time-independent Hamiltonian, which turns out to be the conventional Bose-Hubbard model with the kinetic term renormalized by a gauge field. The same result is obtained upon replacing the classical harmonic oscillation by an auxiliary bosonic quantum field, which is subsequently integrated out. Finally, we discuss the gauge structure underlying the effective Hamiltonian.

\subsubsection{Dyson series}
The time-evolution operator for the one-body Hamiltonian, $H_{\tr{1B}}(t) = H_0 + W(t)$, is $U_{\tr{1B}}(t) = T\{\exp[(-i/\hbar)\int_0^t dt' H_{\tr{1B}}(t')]\}$ where $ T\{\}$ denotes the time-ordering operation.
For times $t$ which are multiples of the revolution time $\tau \equiv 2 \pi/\O$ of the rotor potential (i.e., $t = n \tau$ with some integer $n$), the corresponding Dyson series (up to second order in $-i/\hbar$) is calculated as
\beq\label{Dyson1}
U_{\tr{1B}}(t) &=& 1+\left(\frac{-i}{\hbar} \right) \int_0^{t} ds H_{\tr{1B}}(s)  \nn \\
&+&\left(\frac{-i}{\hbar} \right)^2 \int_0^{t} ds \int_0^{s} ds' H_{\tr{1B}}(s)H_{\tr{1B}}(s') \,\, + \dots  \nn \\
&=& 1- \frac{i}{\hbar} H_0 t - \frac{1}{i \hbar^2 \O} \left( [H_0,Q-Q^\dag] + [Q,Q^\dag] \right) t  \nn \\  \nn \\
&+& \mathcal{O}(\geq \tr{2B}, \geq t) \,\, ,
\eeq
where $\mathcal{O}(\geq \tr{2B}, \geq t)$ denotes terms with at least two-body character scaling with $t$ or higher powers of $t$. Note that each expansion order of the Dyson series proportional to $(-i/\hbar)^n$ with $n\geq2$ can contribute terms linear in $t$, however, each exhibiting at least $n-1$-body character. With $Q$ as defined in Eq.~(\ref{Ham2}) and upon neglecting the $\mathcal{O}(\geq \tr{2B}, \geq t)$ many-body terms, we find
\beq
\label{Dyson2}
U_{\tr{1B}}(t) \!\!&\simeq&\!\! 1- \frac{i}{\hbar} H^{eff}_0\, t \nn\\
H^{eff}_0\!\!&=&\!\! -J_0 \mathcal{T} + \frac{i J_0  \chi_1 }{\hbar\O} [\mathcal{T},\mathcal{M}] - \frac{i \chi_1 \chi_2 }{2\hbar\O} [\mathcal{M},\mathcal{N}]
\eeq
Note that the periodicity $H_{\tr{1B}}(t+\tau) = H_{\tr{1B}}(t)$ implies that $U_{\tr{1B}}(n \tau) = [U_{\tr{1B}}(\tau)]^n$. Thus, it suffices to justify neglecting the many-body terms in the derivation of $H^{eff}_0$ for a single revolution time $\tau$. By using the commutator (anticommutator) relations of the atomic operators according to their bosonic (fermionic) nature, after some algebra we find $[\mathcal{T},\mathcal{M}] = 0$ and
\beq\label{Commutator}
[\mathcal{M},\mathcal{N}] \, = \,2 \sum_{\tb{r} \in \mathcal{A}, l=1-4} (-1)^l \lt\{ a_{\tb{r}}^{\dag} b_{\tb{r}+\tb{e}_l}
-\tr{H.c.} \rt \}\, ,
\eeq
and thus
\beq\label{HamEff}
H^{eff}_0=-J\sum_{\tb{r} \in \mathcal{A}, l=1-4} \lt\{ e^{i\p(-1)^l/4}a_{\tb{r}}^{\dag} b_{\tb{r}+\tb{e}_l}+\tr{H.c.}\rt\},
\eeq
where $J\equiv \sqrt{J_0^2+W_0^2}$, $\p \equiv 4 \tan^{-1}(W_0/J_0)$ and $W_0\equiv  \chi_1\chi_2/\hb \O$. As shown, within an effective time-independent description, the temporal modulation $W(t)$ renormalizes the real isotropic hopping amplitudes $J_0$ of the conventional Hubbard model by adding an anisotropic imaginary contribution. Note that the value of the phase $\p$ in Eq.~(\ref{HamEff}) is limited to the interval $[-2\pi, 2\pi]$, since only $J_0>0$ can be accessed with the rotor technique. As discussed below, the effective Hamiltonian~(\ref{HamEff}) mimics the action upon charged particles of a staggered magnetic field alternating in sign for adjacent plaquettes. A similar effective description has also been used to derive a uniform artificial magnetic field in an optical lattice in Ref.~\cite{Sorensen:05}.

\subsubsection{Auxiliary field method}
The harmonic time dependence in the temporal modulation $W(t) = Q^{\dag} e^{i\O t} + Q e^{-i\O t}$ in Hamiltonian~(\ref{Ham2}) suggests a quantization procedure that allows the time dependence of the system to be eliminated. This method, which is similar to the ``adiabatic elimination" of the excited state of a two-level atom coupled to an off-resonant light field \cite{Marte:93}, amounts to
replacing the classical oscillation terms $e^{\pm i\O t}$ with auxiliary creation (annihilation) operators $\hat{p}^\dag$ ($\hat{p}$), which obey bosonic commutation relations:
\beq
e^{-i\O t} &\rightarrow& \hat{p} ,\nn\\
e^{i\O t} &\rightarrow& \hat{p}^{\dag}.
\eeq
The one-body part $H_{\tr{1B}}(t) = H_0 + W(t)$ of the Hamiltonian~(\ref{Ham2}) thus becomes
\beq\label{Ham7}
H_{\tr{1B}} = -J_0 \mathcal{T} + Q^{\dag} \hat{p}^\dag + \hat{p}\, Q  \, .
\eeq
Note that the replacements are carried out such that the resulting Hamiltonian is written in an inherently Hermitian form. The Heisenberg equation for $\hat{p}$ then reads $i\hb \f{d}{dt}\hat{p} = [\hat{p},H_{1B}] = Q^\dag $, where $[\hat{p},\hat{p}^\dag] = 1$ has been used. Assuming that the evolution of the auxiliary field is entirely determined by external driving, thus neglecting any back action of the atoms upon the rotor potential, leads us to write $\f{d}{dt}\hat{p} = - i \O \, \hat{p}$ and thus $\hat{p} = Q^\dag/ \hb \O$. The latter may be reintroduced into the Hamiltonian~(\ref{Ham7}), yielding
\beq
\label{Ham8}
H_{1B} &=& -J_0 \mathcal{T} +  \frac{2}{\hb \O}    Q^{\dag} Q \nn\\
&=& -J_0 \mathcal{T} - \frac{i \chi_1 \chi_2}{2  \hb \O} [\mathcal{M},\mathcal{N}] \nn\\
&&- \frac{ \chi_1^2 }{2  \hb \O} \mathcal{M}^2- \frac{ \chi_2^2 }{2 \hb \O} \mathcal{N}^2.
\eeq
Comparison with Eq.~(\ref{Dyson2}) shows that the same one-body term is recovered, which gives rise to the staggered magnetic field. The nonlocal two-body terms (proportional to $\mathcal{N}^2$ and $\mathcal{M}^2$) are artifacts resulting from our inappropriate implicit assumption that all atoms interact with the same quantized mode, a scenario not met in experiments, where the rotor potential is essentially classical. Alternatively, the time dependence can be eliminated by means of a path integral method, where the  operators $\hat{p}$ and $\hat{p}^\dag$ are treated as $c$-numbers in a coherent-state representation. The Lagrangian associated with the Hamiltonian~(\ref{Ham7}) contains terms up to quadratic order in the auxiliary quantum field and we may thus integrate them out exactly to arrive at the same effective Hamiltonian~(\ref{Ham8}) \cite{Neg:88}.

\subsubsection{Staggered flux}
A particularly intuitive illustration of the structure of the effective one-body Hamiltonian~(\ref{HamEff}) is obtained if the lattice is composed of plaquettes translated by the primitive vectors $\tb{d}_1$ and $\tb{d}_2$ of the $\mathcal{A}$ sublattice [see Fig.~\ref{Fig.1}(c)]:
\beq
\label{HamP}
H^{eff}_0&=&-J \sum_{\lozenge} e^{i\p/4} \bigl( a^\dag_1 b_2+b^\dag_2 a_3+a^\dag_3 b_4+b^\dag_4 a_1\bigr)\nn\\
&&+ \tr{H.c.}
\eeq
Indices $1-4$ indicate the four corners of a plaquette numbered in clockwise order, starting with the lower left corner. This representation immediately points out that a particle hopping around an elementary plaquette picks up an Aharonov-Bohm phase $\p$ with a sign alternating across adjacent plaquettes, which is equivalent to the presence of a staggered flux with strength $\p$ (in units of the fundamental flux quantum) in each plaquette. For $\p = \pm 2\pi$ we have one flux quantum per plaquette. Notice that to realize this situation for condensed matter, lattice electrons would require unrealistically large magnetic fields in the $10^2-10^3$ Tesla range.

The time-modulation technique used to derive Hamiltonian~(\ref{HamP}) lets us only access fluxes $\p$ in the interval $[-2\pi,2\pi]$ because $J_0 > 0$. Note, however, that Hamiltonian~(\ref{HamP}) displays an $8 \pi$ periodicity with respect to $\p$, which reflects the existence of a second inequivalent flux domain for $\p \in[-4\pi,-2\pi]\cup[2\pi,4\pi]$. This domain corresponds to negative values of $J_0$. A $4 \pi$ change of $\p$, connecting the two domains, reverses the sign of the Hamiltonian. The $8 \pi$ periodicity with respect to $\p$ is in contrast to the case of a uniform magnetic field in a lattice, where the flux per plaquette is defined up to an integer multiple of $2\pi$. The staggered flux in general breaks time-reversal and inversion symmetries, except for the cases of $\p = 2\pi n$ with $n \in \mathbb{Z}$, where the hopping amplitudes attain real ($n=$ even) or imaginary ($n=$ odd) values.

The complex hopping amplitudes are gauge-dependent parameters, whereas the total flux passing through a closed path is gauge-invariant. Recall that an arbitrary lattice Hamiltonian
\beq
H = \sum_{<i,j>}\chi_{ij}c_i^\dag c_j+ \tr{H.c.},
\eeq
with complex nearest-neighbor hopping amplitudes obeying $\chi_{ji}=\chi_{ij}^*$ is invariant under the \textit{local} $U(1)$ gauge transformation
\beq\label{gauge}
c_i&\rightarrow& c_i \exp[-i\th_i],\nn\\
\chi_{ij}&\rightarrow& \chi_{ij}\exp[i (\th_j-\th_i)],
\eeq
It is interesting to note that by means of a gauge change in Eq.~(\ref{HamP}), one can obtain a new Hamiltonian with a periodicity in the flux $\p$ reduced to $2 \pi$: on the plaquette at position $\tb{d}_1 n+ \tb{d}_2 m$, $n,m \in \mathbb{Z}$ the replacements $a_{\nu}\, \rightarrow\, a_{\nu}\, e^{-i \p (\nu + 2n + 2m)/4}$ are made, where $\nu \in \{1,2,3,4\}$ and the $\mathcal{A}$ and $\mathcal{B}$ operators are not explicitly distinguished here. The resulting gauge-transformed Hamiltonian
\beq
\label{HamP2}
H^{eff}_0&=&-J \sum_{\lozenge} \bigl( a^\dag_1 b_2+b^\dag_2 a_3+a^\dag_3 b_4 + e^{i\p} b^\dag_4 a_1\bigr) \nn \\
&&+\tr{H.c.}
\eeq
trivially exhibits $2\pi$ periodicity with regard to $\p$.

To describe a finite system, one conveniently chooses a periodic boundary condition where the opposite sides of the $\sqrt{N}\times \sqrt{N}$ square lattice are identified. This choice results in a topology of a torus for the system considered. If we now compare the total flux gained around a non-contractable loop on the torus, the two Hamiltonians in Eqs.~(\ref{HamP}) and (\ref{HamP2}) give rise to distinct physical realizations. Although there is no net global flux (or $\p/4$ flux) gained in the Hamiltonian~(\ref{HamP}) for $N$ even (odd), a nonzero global flux of $\sqrt{N}\p/2$ is accumulated in the $\tb{e}_1$ direction for Hamiltonian~(\ref{HamP2}). Throughout this paper we will work in the original gauge of Hamiltonian~(\ref{HamP}), which gives the desired physical realization.

\section{2D Massless Dirac Fermions with Anisotropy}\label{sec3}
\begin{figure}
\includegraphics[scale=.32, angle=0, origin=c]{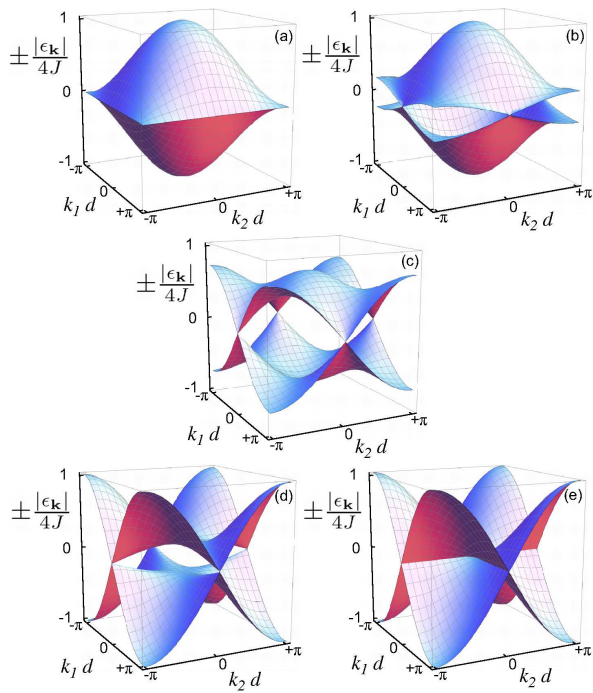}
\caption{\label{Fig.2} (Color online) Single-particle spectra of the ideal lattice fermions subjected to different staggered fluxes: (a) $\p=0$,\,(b) $\p=\pi/4$,\,(c) $\p=\pi$,\,(d) $\p=7\pi/4$,\, (e) $\p=2 \pi$.}
\end{figure}

We now consider loading the optical potential with single-component fermionic atoms. In this case, $s$-wave scattering of the atoms is absent due to the Pauli principle. Furthermore, for the low temperatures considered here, higher angular momentum collision channels are negligible. The effective Hamiltonian in Eq.~(\ref{HamEff}) thus provides a complete description of the system in the tight-binding limit, realizing an ideal lattice Fermi gas in the presence of a staggered flux $\p$. By Fourier transforming the operators
\beq
a_{\tb{r}}&=&\f{1}{\sqrt{N_A}}\sum_{\tb{k}\in1BZ} a_{\tb{k}}e^{i \tb{k}\cdot \tb{r}},\nn\\
b_{\tb{r}+\tb{e}_l}&=&\f{1}{\sqrt{N_B}}\sum_{\tb{k}\in1BZ} b_{\tb{k}}e^{i \tb{k}\cdot (\tb{r}+\tb{e}_l)},
\eeq
Hamiltonian~(\ref{HamEff}) is expressed in momentum space by
\beq\label{Ham6}
H_0^{eff}=-\sum_{\tb{k}\in 1BZ}\e_{\tb{k}}^*a_{\tb{k}}^{\dag}b_{\tb{k}}+\tr{H.c.},
\eeq
with
\beq
\e_{\tb{k}}&=&4J[ \cos (\p/4) \cos(k_1 d/2) \cos(k_2 d/2) \nn\\
&&-i \sin(\p/4) \sin(k_1 d/2) \sin(k_2 d/2)],
\eeq
and the lattice momentum summation is restricted to the first Brillouin zone ($1BZ$) with $k_\nu \equiv \tb{k}\cdot\tb{d}_\nu/d \in [-\pi/d,\pi/d\,], \nu \in \{1,2\}$. The total number of lattice sites is $N=2N_A=2N_B$ with $N_A$ and $N_B$ denoting the number of $\mathcal{A}$- and $\mathcal{B}$-sites, respectively. Upon performing the canonical transformation
\beq
\label{con}
a_{\tb{k}}=\f{1}{\sqrt{2}}\f{\e_{\tb{k}}^*}{|\e_{\tb{k}}|}(-\A_{\tb{k}}+\B_{\tb{k}}),\tr{\ \ }b_{\tb{k}}=\f{1}{\sqrt{2}}(\A_{\tb{k}}+\B_{\tb{k}}),
\eeq
Hamiltonian (\ref{Ham6}) becomes
\beq\label{Ham4}
H_0^{eff}=\sum_{{\tb{k}}\in1BZ} \biggl(-|\e_{\tb{k}}|\B_{\tb{k}}^\dag \B_{\tb{k}}+|\e_{\tb{k}}|\A^\dag_{\tb{k}} \A_{\tb{k}}\biggr),
\eeq
where the single-particle spectrum is given by
\beq
\label{spectrum}
|\e_{\tb{k}}|&=&2J[\cos^2(k^+d)+\cos^2(k^-d)\nn\\
&&+2 \cos(\p /2)\cos(k^+d)\cos(k^- d)]^{1/2}
\eeq
with $k^{\pm}\equiv (k_1\pm k_2)/2$. The energy spectrum (shown in Fig.~\ref{Fig.2} for different flux values $\p$) consists of an upper and a lower band due to the bipartite lattice structure. The operators $\A^\dag_{\tb{k}}$ and $\B^\dag_{\tb{k}}$ create a quasiparticle in the upper band with energy $|\e_{\tb{k}}|$ and in the lower band with energy $-|\e_{\tb{k}}|$, respectively. Note that $\e_{\tb{k}}$ shares the $8\pi$ periodicity of Hamiltonian~(\ref{HamP}) with respect to $\p$, whereas the energy spectrum $|\e_{\tb{k}}|$ exhibits a $4\pi$ periodicity.

For zero flux $\p=0$, the upper and lower energy bands recombine at the Brillouin zone edges. By mapping the upper energy band to the second Brillouin zone, we recover the standard tight-binding energy dispersion in the absence of a gauge field, with the unit cell consisting of an elementary plaquette (the new Brillouin zone is then rotated by $\pi/4$ and expanded by a factor of $\ro{2}$). The presence of the staggered flux immediately leads to interesting properties in the energy band structure. For $\p\neq 2 n \pi, n \in \mathbb{Z}$, the upper and lower energy bands intersect at four conical points $(\pm \pi/d,0)$ and $(0,\pm\pi/d)$ on the Brillouin zone edges. However, there are only two inequivalent points, which we denote $K_+=(\pi/d,0)$ and $K_-=(0,\pi/d)$, given by the zeros of the energy spectrum $|\e_{K_{\pm}}|=0$. At half-filling, the lower energy band is completely filled and the Fermi level coincides with the conical points giving rise to exact particle-hole symmetry. An expansion of the energy dispersion for small momenta around either of the conical points $K_{\pm}$ gives
\beq
|\e_{K_{\pm}+\tb{k}}|&=&\sqrt{2}J d\biggl\{\left[1\pm\cos(\p/2)\right]  k_1^2\nn\\
&&+\left[1\mp\cos(\p/2)\right]  k_2^2\biggr\}^{1/2}+{\cal O} \lt(|\tb{k}|^2\rt).
\eeq
We see that the low-energy excitations disperse linearly in momentum, (i.e., they are Dirac-like) in contrast to the case of ordinary particles with a quadratic dispersion. By defining the Fermi velocity $\hb v_{\tr{F}}=\sqrt{2}Jd$, the low-energy Hamiltonian becomes
\beq
H_0^{eff}\!\!\!&\simeq&\!\!\!\sqrt{2}\hb v_F\!\!\!\sum_{\tb{k}\in 1BZ}\biggl\{\left[\cos(\p/4)  k_1- i\sin(\p/4)  k_2 \right]a_{+,\tb{k}}^{\dag}b_{+,\tb{k}}\nn\\
&&\!\!\!\!\!\!\!\!+\left[\cos(\p/4)  k_2-i\sin(\p/4)  k_1 \right]a_{-,\tb{k}}^{\dag}\h{b}_{-,\tb{k}}+\tr{H.c.}\biggr\},
\eeq
which contains two copies of Dirac-like particles described by the operators $(a_{+,\tb{k}},b_{+,\tb{k}})$ and $(a_{-,\tb{k}},b_{-,\tb{k}})$, one around each individual Dirac point $K_{\pm}$, respectively. Notice that several remarkable phenomena, for example, the Klein paradox and the phenomenon of \textit{Zitterbewegung}, expected for noninteracting Dirac particles in two dimensions, are to be met here. We refer the interested reader to the review work about graphene, the prototypical system exhibiting Dirac electrons, in Ref.~\cite{Neto:09}.

The Dirac cones arising here are generally anisotropic (cf. Fig.~\ref{Fig.2}), which results in anisotropic propagation velocities. The anisotropy of the cone is controlled by the staggered flux. Only at the special value $\p=\pi$ do the Dirac cones become isotropic. At this point, the system simulates the mean-field Hamiltonian of the $\pi$-flux phase proposed by Affleck and Marston to describe the pseudogap regime of the high-$\tr{T}_c$ cuprates \cite{Affleck:88}. Furthermore, the picture also becomes reminiscent of graphene tight-binding physics.

The adjustable anisotropy of the Dirac cones is a specific feature of the staggered-flux scenario in a square lattice and is intimately connected to the breaking of time-reversal and inversion symmetries. It does not arise in graphene or graphene-like system with cold atoms in a hexagonal optical lattice \cite{Zhu:07}. For graphene, the insertion of a time-reversal symmetry breaking perturbation would move the two inequivalent Dirac points towards each other, or produce a gap in the spectrum, while the isotropy of the cones is maintained \cite{Manes-Vozmediano07}. According to Ref.~\cite{ParkNature08}, anisotropic Dirac cones could be engineered in graphene by growing the graphene on top of a suitably patterned periodic potential. However, the anisotropy would then be fixed. The \textit{in situ} tuning of the cone anisotropy in our system is reminiscent of options arising in organic compounds (see Ref. \cite{Goerbig:08}), which provide Dirac cones with a tunable tilt.

In summary, by loading the staggered optical lattice with single-component fermions, we obtain an ideal Dirac system with tunable anisotropic Dirac cones at half-filling. The next section discusses the case for bosons, where interactions become important.

\section{Bosonic Superfluid States}\label{sec4}
For single-component bosons, the non-vanishing $s$-wave collisions between the atoms give rise to the onsite Hubbard interaction $H_{\tr{int}}$ of Eq.~(\ref{Ham2}). The staggered flux modifies the hopping term of the conventional Hubbard model according to Eq.~(\ref{HamEff}). Therefore, we now study the generalized Bose-Hubbard model
\beq
\label{Ham5}
H_{BH}=H_0^{eff}+\f{1}{2}U\sum_{\tb{r}\in \mathcal{A}\oplus \mathcal{B}}n_{\tb{r}}\,\,(n_{\tb{r}}-1),
\eeq
with $H_0^{eff}$ given by the Hamiltonian~(\ref{HamEff}). This section considers the weakly interacting regime governed by the physics of Bose-Einstein condensation. It is shown that, for different flux values $\p$, distinct superfluid phases can be realized: a homogeneous zero-momentum superfluid for $-\pi< \p < \pi$; a finite-momentum superfluid for $-3\pi< \p < - \pi$ or  $\pi< \p < 3\pi$, characterized by a vortex-antivortex lattice with one vortex per plaquette and different rotational directions for the two flux intervals; and finally, for $-4\pi< \p <- 3\pi$ and $3\pi< \p < 4\pi$, a finite-momentum superfluid with an order parameter that has opposite sign for adjacent lattice sites.

For sufficiently weak interactions and low temperatures, the atoms Bose-condense in the lowest-energy single-particle state. The many-body ground state is then well described by the Hartree expression $\dr{\Psi_{\tb{k}_0}}=(\B_{\tb{k}_0}^\dag)^{N_0}\dr{0}$, where $N_0$ is the number of condensed atoms, $\tb{k}_0$ is the quasimomentum of the lowest-energy single-particle state, and $\B_{\tb{k}_0}^\dag$ is the corresponding quasiparticle creation operator introduced in Eq.~(\ref{con}).

\begin{figure}
\includegraphics[scale=.19, angle=0, origin=c]{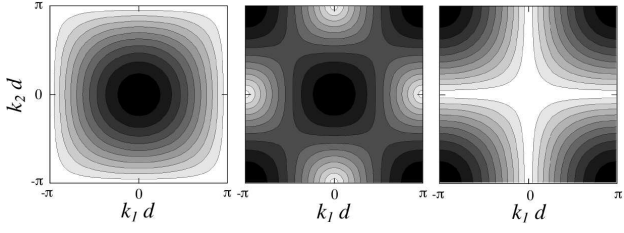}
\caption{\label{Fig.3}
Contour plots of the single-particle spectra for different staggered fluxes. Dark regions indicate low energy. (a) For $\p=n\,4\pi$, $n$ integer, the minimum occurs at $\tb{k}=(0,0)$. (b) For $\p=\pi+n\,2\pi$, $n$ integer, degenerate minima occur at $(0,0)$ and $(\pm\pi/d,\pm\pi/d)$. (c) For $\p=2\pi+n\,4\pi$, $n$ integer,  equivalent minima occur at $(\pm\pi/d,\pm\pi/d)$.}
\end{figure}

As illustrated in Fig.~\ref{Fig.3}, the minima of the lower band of the single-particle spectrum $-|\e_{\tb{k}}|$ in Eq.~(\ref{spectrum}) arise at positions in $k$-space depending on the value of the flux $\p$. Two distinct cases arise: the lowest-energy state occurs at the center of the Brillouin zone $\tb{k}_0=(0,0)\equiv 0$ if $-\pi<\p + 4\pi m_{0}< \pi$, or it occurs at the four corners of the Brillouin zone $\tb{k}_0=(\pm\pi/d,\pm\pi/d) \equiv \pi$ if $\pi<\p + 4\pi m_{\pi}< 3\pi$. Here, $m_{0}$ and $m_{\pi}$ are arbitrary integers. In the $\tb{k}_0=0$ case, using $\e_{\tb{k}=0}^*/|\e_{\tb{k}=0}| = {\rm sgn }[\cos(\phi/4)]$ in Eq.~(\ref{con}), we may calculate $\dr{\Psi_{\tb{k}_0}}$ in configuration space,
\beq\label{Har1}
\dr{\Psi_{0,(-1)^{m_{0}}}}= \qquad\qquad\qquad\qquad\qquad \nn\\
\left\{\f{1}{\sqrt{N}} \sum_{\tb{r}\in \mathcal{A}} \left( (-1)^{m_{0}} a_{\tb{r}}^\dag +b_{\tb{r}+e_1}^\dag \right) \right\}^{N_0}\dr{0}.
\eeq
Depending upon whether $m_{0}$ is even or odd, we obtain different superfluid phases. For even $m_{0}$, which corresponds to positive values of the hopping strength $J_0$, we recover the familiar zero-momentum homogeneous superfluid state known from the conventional Bose-Hubbard model. For odd $m_{0}$, the boson operators occur with different signs for the two sublattices $\mathcal{A}$ and $\mathcal{B}$; that is, the order parameter is constant except for a different sign at the $\mathcal{A}$ and $\mathcal{B}$-sites (referred to as staggered--sign superfluid).

To understand better how the staggered-sign superfluid arises, note that although the energy band structure (in Fig.~\ref{Fig.2}) remains invariant under the transformation
\beq
a_{\tb{k}}\rightarrow a_{\tb{k}}, \tr{\ \ \ \ \ }
b_{\tb{k}}\rightarrow -b_{\tb{k}},
\eeq
the corresponding upper- and lower-band states are interchanged. The staggered-sign superfluid and the uniform superfluid are thus distinct, despite the fact that they arise for the same lattice momentum $\tb{k}=0$.

In the case $\tb{k}_0 = \pi$, each of the four equivalent minima at the corners of the Brillouin zone, which are related to each other by reciprocal lattice vectors, yields $\e_{\tb{k}=\pi}^*/|\e_{\tb{k}=\pi}| = i\,{\rm sgn }[\sin(\phi/4)]$. Introducing this into Eq.~(\ref{con}) leads to the $k$-space expression
\beq\label{staggered}
\dr{\Psi_{\pi,(-1)^{m_{\pi}}}}&=&  \left\{\f{1}{\sqrt{2}} \left((-1)^{m_{\pi}} i \,a_{\pi}^{\dag}+b_{\pi}^{\dag}\right)\right\}^{N_0} \dr{0}. \nn \\
\eeq
After Fourier-transforming the creation operators $a^\dag_{\pi}=N_A^{-1/2} \sum_{m,n\in\mathbb{Z}} \,\, a_{(m,n)}^{\dag} e^{i \pi (m+n)}$ and similarly $b^\dag_{\pi}= N_B^{-1/2}\sum_{m,n\in\mathbb{Z}}\,\, b_{(m,n)+\tb{e}_1}^{\dag} e^{i \pi (m+n+1)}$ with $(m, n)\equiv m \,\tb{d}_1+n\, \tb{d}_2$, we obtain the ground-state wave function (\ref{staggered}) in real space
\beq
\label{Har2}
&&\dr{\Psi_{\pi,(-1)^{m_{\pi}}}} =  \,\,\, \left(\f{i (-1)^{m_{\pi}}}{\sqrt{N}}\right)^{N_0}  \\
&&\times\lt\{\sum_{\square} \biggl( a_{1}^\dag + (-1)^{m_{\pi}} i b_{2}^\dag - a_{3}^\dag - (-1)^{m_{\pi}} i b_{4}^\dag\biggr) \rt\}^{N_0}\dr{0}, \nn
\eeq
where $\sum_{\square}$ denotes the summation over the \textit{shaded} plaquettes shown in Fig.~\ref{Fig.1}(d). One recognizes that this wave function (referred to as staggered--vortex superfluid) accumulates a phase of $\pm 2\pi$, when moving around an elementary plaquette, with alternating sign for adjacent plaquettes. This forms a lattice of singly quantized staggered vortices, which are commensurate with the external staggered flux. The Bose-Einstein condensate (BEC) formed for the magnetic flux $\pi<\p + 4\pi m_{\pi} < 3\pi$ is thus characterized by a vortex-antivortex lattice, whereas the rotational direction on a given plaquette is determined upon whether $m_{\pi}$ is even or odd.

Next, we consider the stability of the two BECs which can arise for flux values in the interval $[-2\pi,2\pi]$,
given by the states $\dr{\Psi_{0,(-1)^{m_{0}}}}$ and $\dr{\Psi_{\pi,(-1)^{m_{\pi}}}}$ in Eqs.~(\ref{Har1}) and (\ref{Har2}) for both $m_0$ and $m_{\pi}$ even. We employ a variational approach for the BEC ground state with the ansatz
\beq
\dr{\xi,\s}\!=\!\f{1}{\sqrt{N_0}}(e^{-i\xi/2}\cos(\s)\B_0^\dag
+e^{i\xi/2}\sin(\s)\B_\pi^\dag)^{N_0}\dr{0},
\eeq
where the two variational parameters $\xi$ and $\s$ are to be determined by minimizing the ground-state energy at zero temperature. With respect to the Hamiltonian~(\ref{Ham5}), the variational ground-state energy is calculated to be
\beq
\label{MF}
&&\langle H_{BH}\rangle=-4N_0J\sin\biggl(\f{\phi}{4}\biggr)-\f{UN_0(N_0-1)}{N}\cos^4(\s)\nn\\
&&+\biggl\{ 4J\biggl[\sin\biggl(\f{\phi}{4}\biggr)-\cos\biggl(\f{\phi}{4}\biggr)\biggr]+\f{U (N_0-1)}{ N}\biggr\}N_0\cos^2(\s)\nn\\
&&+\f{U N_0(N_0-1)}{2N}\equiv E_{MF}.
\eeq
The first observation is that the $\xi$ dependence drops out completely in the mean-field energy $E_{MF}$. This can be understood at the variational level from the fact that the Hamiltonian is not sensitive to the relative phase difference between the condensation points. Next, we see that for $0\leqslant\p<\pi$ ($\pi<\phi\leqslant 2\pi$) the $\tb{k}=0$ uniform BEC with $\s_0=0$ (the $\tb{k}=\pi$ staggered-vortex BEC with $\s_0=\pi/2$) is indeed the absolute minimum of the mean-field energy. Finally, the stability of the different ground states is verified by allowing a small deviation $\varepsilon$ from the condensation point $\tb{k}_0$. The variation in energy is then given by
\beq
\langle H_{BH}\rangle_{\s_0+\varepsilon}&=&E_{MF}+4 \varepsilon^2  N_0 J\biggl[ \ro{2} \sin\biggl(\biggl|\f{\p-\pi}{4}\biggr|\biggr)\nn\\
&&+\f{N_0U}{4JN}\biggr]+O(\varepsilon^4).
\eeq
Since the quantity in the bracket is positive definite, we conclude that the ground state is stable against small variations.

For $\phi=\pi$, the mean-field energy exhibits two degenerate minima at the two points $\s_0=0$ and $\s_0=\pi/2$ separated by an energy barrier $\sim UN_0^2/4N$. The absence of a $\xi$ dependence in the mean-field energy precludes a coherent superposition state of the two condensation points at the flux value $\phi=\pi$. It thus suggests that the two superfluid phases are separated by a first-order quantum phase transition, where the order parameter changes discontinuously across this point. We remark that the Bose-Hubbard model with $\phi=\pi$ is equivalent to the fully frustrated Josephson junction model \cite{Polini:05}.

Having shown the stability of the distinct BEC ground states, we now study their excitation spectrum using Bogoliubov theory. We first write Hamiltonian (\ref{Ham5}) in the grand canonical ensemble by introducing a chemical potential $\mu$:
\beq
H_0^{eff}-\mu N= \,\,\,  \qquad \qquad \qquad \qquad \qquad \qquad \qquad \qquad  \\
\sum_{\tb{k}\in1BZ}\biggl[ (-|\e_{\tb{k}}|-\mu)\B_{\tb{k}}^\dag \B_{\tb{k}}+(|\e_{\tb{k}}|-\mu)\A^\dag_{\tb{k}} \A_{\tb{k}}\biggr], \nn
\eeq
and for the interactions
\beq
H_{int}=\f{U}{N}\sum_{\substack{\tb{k}_1+\tb{k}_2\\=\tb{k}_3+\tb{k}_4}}\biggl[a^\dag_{\tb{k}_1} a^\dag_{\tb{k}_2} a_{\tb{k}_3}a_{\tb{k}_4} +b^\dag_{\tb{k}_1} b^\dag_{\tb{k}_2} b_{\tb{k}_3}b_{\tb{k}_4}\biggr]
\eeq
we perform the canonical transformation (\ref{con}). By identifying the condensation mode $\B_{k_0}\rightarrow\ro{N_0}+\B_{k_0}$, we perform the Bogoliubov approximation while keeping the fluctuation modes only up to quadratic order. The chemical potential is chosen such that the terms which are linear in the fluctuation vanish; that is,
\beq
\mu=-|\e_{k_0}|+n_0U,
\eeq
where $n_0=N_0/N$ is the condensate density. After some algebra, we obtain the action for the fluctuations
\beq
S[\Phi,\Phi^\dag]\approx-\f{1}{2}n_0UN_0-\f{\hb}{2}\sum_{\tb{k},m}\Phi_{\tb{k}}^\dag\cdot \tb{G}_{\tb{k}}^{-1}\cdot \Phi_{\tb{k}},
\eeq
where $\Phi^\dag_{\tb{k}}\equiv (\A_{\tb{k}}^*,\A_{-\tb{k}},\B_{\tb{k}}^*,\B_{-\tb{k}}) $ are the fluctuation fields and the one-particle Green function $\tb{G}_{\tb{k}}$ in the Nambu space is given by
\begin{widetext}
\beq
-\hb \tb{G}_{\tb{k}}^{-1}\equiv\bem
-i\hb \o_m+|\e_{\tb{k}}|+M_{\tb{k}_0}  & \frac{1}{2}n_0U
A_{\mathbf{k},\mathbf{k}_0}&0&\frac{1}{2}n_0U
B_{\mathbf{k},\mathbf{k}_0}\\
 \frac{1}{2}n_0U
A_{\mathbf{k},\mathbf{k}_0}^*&i\hb \o_m+|\e_{\tb{k}}|+M_{\tb{k}_0} &\frac{1}{2}n_0U
B_{\mathbf{k},\mathbf{k}_0}^*&0\\
0&\frac{1}{2}n_0U
B_{\mathbf{k},\mathbf{k}_0}&-i\hb \o_m-|\e_{\tb{k}}|+M_{\tb{k}_0}& \frac{1}{2}n_0U
A_{\mathbf{k},\mathbf{k}_0}\\
\frac{1}{2}n_0U
B_{\mathbf{k},\mathbf{k}_0}^*&0&\frac{1}{2}n_0U
A_{\mathbf{k},\mathbf{k}_0}^*&i\hb \o_m-|\e_{\tb{k}}|+M_{\tb{k}_0}\eem\nn
\eeq
\end{widetext}
where $M_{\tb{k}_0}=|\e_{\tb{k}_0}|+n_0 U$, $A_{\mathbf{k},0}=B_{\mathbf{k},\pi}=1+\exp(-2i\varphi_\mathbf{k})$, $B_{\mathbf{k},0}=A_{\mathbf{k},\pi}=1-\exp(-2i\varphi_\mathbf{k})$, and $\varphi_\mathbf{k}=\arg (\epsilon_{\mathbf{k}})$.
To obtain the excitation spectrum, we go back from the Matsubara frequency to real time $i\hb \o_m\rightarrow \hb\o$ and find the poles of the $4\times 4$ one-particle Green function. This can be easily done by determining the eigenfrequency of the equation $\tr{det}[-\hb \tb{G}_{\tb{k}}^{-1}]=0$, which yields
\beq
\hb\o=\sqrt{|\epsilon_\mathbf{k}|^2+|\epsilon_{\mathbf{k}_0}|^2+2 n_0U|\epsilon_{\mathbf{k}_0}|\pm 2n_0U|\epsilon_\mathbf{k}| \sqrt{F_{\mathbf{k},\mathbf{k}_0}}},\nn
\eeq
where $F_{\mathbf{k},\mathbf{k}_0}=\cos^2 ( \varphi_\mathbf{k} )+2 |\epsilon_0|[|\epsilon_0|/2n_0U+1]/n_0U$. Once again, the two branches of the excitation spectrum are due to the sublattice degrees of freedom. To examine the long wavelength modes, we perform a Taylor expansion around the condensation momentum $k_0$ in the lower branch and get $E_k\approx v (\mathbf{k}-\mathbf{k}_0)$, with the speed of sound
\beq
v=\sqrt{J\cos\biggl(\f{\p}{4}-\f{k_0 d}{2}\biggr)\biggl[4J\cos\biggl(\f{\p}{4}-\f{k_0 d}{2}\biggr) +2n_0 U\biggr]},\nn
\eeq
corresponding to the Goldstone mode of the broken gauge symmetry.

\section{Superfluid-Mott Insulating Transition}\label{sec5}
In this section, we determine the complete phase diagram of the generalized Bose-Hubbard model in the strong coupling regime at zero temperature. In the absence of the external staggered gauge field, the zero-temperature phase diagram of the Bose-Hubbard model comprises a superfluid (SF) phase and a Mott insulator (MI) phase. These phases are separated by a second-order phase transition, driven by quantum fluctuations, which is controlled by the dimensionless number $U/4J_0$. When crossing the phase boundary into the SF phase, the $U(1)$ gauge symmetry is spontaneously broken, thus giving rise to an SF-order parameter. In Sec. \ref{sec4} it was shown that, in the presence of the staggered flux $\p$, the broken-symmetry phase consists of distinct SF phases. As the interaction strength is increased, we expect a SF-MI transition to take place for each of these SF phases. We first use Landau's theory of phase transitions by introducing a plaquette order parameter, which takes into account the various SF phases. Within this framework, we determine the critical coupling strength $(U/4J)_c$, where the SF order is destroyed. Next, we study the Mott regime in detail and derive the excitation spectrum using the path integral formalism. In contrast to the Landau theory, we introduce a Hubbard-Stratonovich field in the Mott regime to characterize the Mott state and treat the hopping terms as perturbations.

\subsection{Landau Theory of Phase Transitions}
For convenience, in this subsection we write the Hamiltonian (\ref{Ham5}) in the plaquette notation of Fig.~\ref{Fig.1}(c). The creation (annihilation) operators $a_\nu^\dag (a_\nu)$, with $\nu=1,2,3,4$, are labelled according to the four sites of an elementary plaquette without explicitly distinguishing $\mathcal{A}$ and $\mathcal{B}$ operators. The Hamiltonian then becomes
\beq
H&=&\sum_{\lozenge}\biggl\{ -J e^{i\p/4} \bigl( a^\dag_1 a_2+a^\dag_2 a_3+a^\dag_3 a_4+a^\dag_4 a_1\bigr)+\tr{H.c.}\nn\\&&\tr{\ \ \ \ \ \ \ }+\f{U}{4}\sum_{\nu=1}^4 n_\nu (n_\nu-1) \biggr\},
\eeq
where $n_\nu=a_\nu^\dag a_\nu$ and $\sum_{\lozenge}$ is the summation over the \textit{shaded} plaquettes, as shown in Fig.~\ref{Fig.1}(c). We anticipate broken-symmetry SF phases to emerge for weak interactions and introduce a plaquette order parameter $\psi \equiv (\psi_1,\psi_2,\psi_3,\psi_4)$ to characterize them. By performing a mean-field decoupling in the hopping term
\beq
a_\nu^\dag a_{\nu'}&=& (\psi_\nu^*+a_\nu^\dag-\psi_\nu^*) (\psi_{\nu'}+a_{\nu'}-\psi_{\nu'})\nn\\
&\simeq&\psi_\nu^* a_{\nu'}+a_\nu^\dag \psi_{\nu'}-\psi^*_\nu \psi_{\nu'},
\eeq
with $\nu,\nu' \in \{1,2,3,4\}$, we find the mean-field Hamiltonian $H_{0,MF}+H_{1,MF}$ in the grand canonical ensemble
\beq
H_{0,MF}&=&\sum_{\lozenge}\sum_{\nu=1}^4 \biggl(\f{U}{4}n_\nu (n_\nu-1)-\f{\mu}{2}n_\nu \nn \\
&&+Je^{-i\p/4}\psi_\nu^*\psi_{\nu+1}
+Je^{i\p/4}\psi_\nu\psi_{\nu+1}^*\biggr), \nn
\eeq
\beq
H_{1,MF}&=&4J\sum_{\lozenge}\biggl[(e^{i\p/4}\psi_1+e^{-i\p/4}\psi_3)a_2^\dag\nn\\&&+(e^{-i\p/4}\psi_1+e^{i\p/4}\psi_3)a_4^\dag \nn \\
&&+(e^{-i\p/4}\psi_2+e^{i\p/4}\psi_4)a_1^\dag\nn\\&&+(e^{i\p/4}\psi_2+e^{-i\p/4}\psi_4)a_3^\dag\nn+\tr{H.c.}\biggr].
\eeq
We see that $H_{0,MF}$ is diagonal in the number-state basis. This
allows us to calculate the ground-state energy $E[\psi]$ up to the
second order with respect to the perturbation $H_{1,MF}$ to get
\beq
\label{bilinear}
E[\psi] &=& n(n-1) \,\bar{U} - 2\,n\,\bar{\mu} \\ \nn
&+&\sum_{\nu,\nu'}\psi^*_\nu \,
M_{\nu\nu'}(n,\bar{U},\bar{\mu},\p)\,\psi_{\nu'}+\mathcal{O}(\psi^4),
\eeq
where $n$ is the filling fraction and $\bar{U}\equiv U/4J$,
$\bar{\mu}\equiv \mu/4J$ are the dimensionless interaction strength and
chemical potential, respectively. The $4\times 4$ Hermitian matrix
$M_{\nu,\nu'}$ is given by
\begin{widetext}
\beq
M(n,\bar{U},\bar{\mu},\p)=\bem
  E^{(0)}(n,\bar{U},\bar{\mu}) &e^{-i\p/4}&E^{(0)}(n,\bar{U},\bar{\mu})
\cos(\p/2)&e^{i\p/4}\\
e^{i\p/4}&E^{(0)}(n,\bar{U},\bar{\mu})&e^{-i\p/4}&E^{(0)}(n,\bar{U},\bar{\mu})
\cos(\p/2)\\
E^{(0)}(n,\bar{U},\bar{\mu})
\cos(\p/2)&e^{i\p/4}&E^{(0)}(n,\bar{U},\bar{\mu})&e^{-i\p/4}\\
e^{-i\p/4}&E^{(0)}(n,\bar{U},\bar{\mu})\cos(\p/2)&e^{i\p/4}&E^{(0)}(n,\bar{U},\bar{\mu})\eem,
\eeq
\end{widetext}
where
\beq
E^{(0)}(n,\bar{U},\bar{\mu})=\biggl[\f{n}{\bar{U}(n-1)-\bar{\mu}}+\f{n+1}{\bar{\mu}-\bar{U}n}\biggr]\,.
\eeq
In the standard Landau theory, the free energy is expanded with respect to a scalar order parameter and the vanishing of the second-order expansion coefficient determines the second-order phase transition point. In the present extension of Landau's theory, second-order phase transitions occur at the zero crossings of the eigenvalues of the matrix $M(n,\bar{U},\bar{\mu},\p)$ in Eq.~(\ref{bilinear}). There are four eigenvectors and respective eigenvalues of the matrix $M(n,\bar{U},\bar{\mu},\p)$ corresponding to the four possible SF phases found in Eqs.~(\ref{Har1}) and (\ref{Har2}), namely the zero-momentum homogeneous SF and the staggered-sign SF,
\beq
\label{order1}
\psi_{0,\pm}&=&(1,\pm1,1,\pm 1), \\ \nn
\varepsilon_{0,\pm}&=&2 \cos(\p/4)[E^{(0)}(n,\bar{U},\bar{\mu})\cos(\p/4) \pm 1],
\eeq
and the two staggered-vortex SF order parameters with opposite rotational directions,
\beq
\label{order2}
\psi_{\pi,\pm}&=&(1,\pm i,-1,\mp i), \\ \nn
\varepsilon_{\pi,\pm}&=&2 \sin(\p/4)[E^{(0)}(n,\bar{U},\bar{\mu})\sin(\p/4) \pm 1] \,.
\eeq
Zero crossings exist for $\varepsilon_{0,+}$ if $-\pi<\p + 4\pi m_{0}< \pi$ and $m_{0}$ is even, for $\varepsilon_{0,-}$ if $-\pi<\p + 4\pi m_{0}< \pi$ and $m_{0}$ is odd, for $\varepsilon_{\pi,+}$ if $\pi<\p +  4\pi m_{\pi} < 3\pi$ and $m_{\pi}$ is even, and for $\varepsilon_{\pi,-}$ if $\pi<\p + 4\pi m_{\pi}  < 3\pi$ and $m_{\pi}$ is odd. We may thus determine the phase boundaries where there is a phase transition between the SF and the MI in the different regimes of $\p$ as
\beq
\label{zero1}
\bar{\mu}_{0,\pm}&=&\f{1}{2}\biggl[\bar{U}(2n-1)\mp \cos\biggl(\f{\p}{4}\biggr)\biggr] \\  \nn
&& \pm\f{1}{2}\sqrt{\biggl[\bar{U} \mp \cos\biggl(\f{\p}{4}\biggr)\biggr]^2\mp 4n\bar{U} \cos\biggl(\f{\p}{4}\biggr)}
\eeq
and
\beq
\label{zero2}
\bar{\mu}_{\pi,\pm}&=&\f{1}{2}\biggl[\bar{U}(2n-1)\mp\sin\biggl(\f{\p}{4}\biggr)\biggr]  \\ \nn
&& \pm\f{1}{2}\sqrt{\biggl[\bar{U}\mp \sin\biggl(\f{\p}{4}\biggr)\biggr]^2\mp 4n\bar{U} \sin\biggl(\f{\p}{4}\biggr)}.
\eeq
In Fig.~\ref{Fig.4}(a), the surfaces bounding the $n=1$ and $n=2$ Mott lobes, given by Eqs.~(\ref{zero1}) and~(\ref{zero2}), are shown as a function of the experimentally relevant parameters $(U/4J_0,\mu/4J_0,W_0/J_0)$ for the first quadrant of the complex $(J_0 + i W_0)$-plane ($0\leq J_0,W_0$, i.e., $0 \leq \p < 2\pi$). Outside the Mott lobes, the two types of SF orders are separated by the horizontal plane at $W_0=J_0$, corresponding to a flux $\p=\pi$. The plane spanned by the $W_0$ axis and the white dashed line in Fig.~\ref{Fig.4}(a), given by $\mu/U=2-\sqrt{2}$, corresponds to a filling factor of unity. For this plane, the complete phase diagram covering the entire range $[-4\pi,4\pi]$ of $\p$ is plotted in Fig.~\ref{Fig.4}(b) .

\subsection{Effective Action for the Mott State}
\begin{figure}
\includegraphics[scale=.32, angle=0, origin=c]{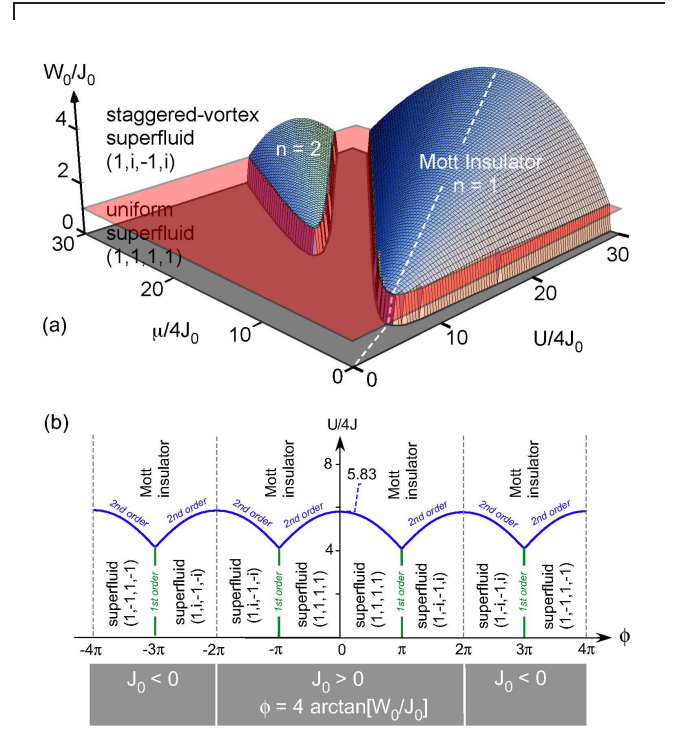}
\caption{\label{Fig.4}(Color online) (a) Phase diagram of the generalized Bose-Hubbard model subjected to a staggered flux $\p$ for the first quadrant of the complex $(J_0 + i W_0)$-plane ($0\leq J_0,W_0$, i.e., $0 \leq \p < 2\pi$). Outside the Mott lobes, two types of superfluid orders arise, separated by the horizontal plane at $W_0=J_0$ corresponding to a flux $\p=\pi$. The plane spanned by the $W_0$ axis and the white dashed line, given by $\mu/U=2-\sqrt{2}$, corresponds to unity filling factor. (b) Phase diagram for unity filling factor covering the entire allowed range $[-4\pi,4\pi]$ of $\p$. The range $\p \in [-2\pi,2\pi]$, corresponding to positive $J_0$, is accessible by the time-modulation technique discussed in Sec.\ref{sec2}, which yields a flux $\p = 4 \arctan[W_0/J_0]$. The superfluid phases are indicated by their plaquette order parameters according to Eqs.~(\ref{order1}) and (\ref{order2}).}
\end{figure}

We now employ a path integral formulation to derive the excitation spectrum of the Mott state in the strong coupling regime, thus generalizing a method presented in Ref.~\cite{Oosten:01}. We first write the partition function for the generalized Bose-Hubbard model in terms of the path integral $Z=\int \mathcal{D}a^* \mathcal{D}a \exp\{ -S[a^*,a]/\hb\}$, where the Euclidean action in the grand canonical ensemble is given by
\beq
&&S[a^*,a]=\int_0^{\hb \B} d\t \biggl[ \sum_{i\in \mathcal{A}\oplus\mathcal{B}} a_i^*(\t)(\hb\partial_\t-\mu)a_i(\t)\nn\\
&&-\sum_{<i,j>}\chi_{ij}a_i^*(\t) a_j(\t)+\f{1}{2}U\sum_{i\in \mathcal{A}\oplus\mathcal{B}} a_i^*(\t)a_i^*(\t)a_i(\t) a_i(\t)\biggr]\nn
\eeq
and the hopping matrix elements $\chi_{ij}$ are given by $\chi_{\tb{r},\tb{r}\pm e_1}=J\exp(i\phi/4)$ and $\chi_{\tb{r},\tb{r}\pm e_2}=J\exp(-i\phi/4)$ where $\tb{r}\in \mathcal{A}$. Since we are interested in the Mott regime where onsite interactions are important, we seek to treat the hopping terms $\sum_{<i,j>}\chi_{ij}a_i^*(\t) a_j(\t)$ as perturbations. This is achieved by introducing the Hubbard-Stratonovich field $(\psi_i(\t),\psi_i^*(\t))$, such that the hopping terms can be decoupled in the following way:
\begin{widetext}
\beq
Z&=&\int  \mathcal{D}\psi^* \mathcal{D}\psi \mathcal{D}a^* \mathcal{D}a \exp\biggl\{ -\f{1}{\hb}\int d\t \sum_{<i,j>} \biggl(\psi^*_i(\t)-a_i^*(\t)\biggr)\chi_{ij}\biggl(\psi_j(\t)-a_j(\t)\biggr) \biggr\} \exp\biggl\{-\f{S[a^*,a]}{\hb} \biggr\}\nn\\
&=&\int \mathcal{D}\psi^* \mathcal{D}\psi\exp\biggl[-\f{1}{\hb} \int d\t\sum_{<i,j>}\psi_i^*(\t)\chi_{ij}\psi_j(\t)\biggr]\nn\\&&\times \int \mathcal{D}a^* \mathcal{D}a \exp\biggl[-\f{1}{\hb}\int d\t \sum_{<i,j>}\biggl(-\psi_i^*(\t) \chi_{ij}a_j(\t)-a_i^*(\t) \chi_{ij}\psi_j(\t)\biggr) \biggr]\exp \biggl\{ -\f{\bar{S}_0[a^*,a]}{\hb}\biggr\}\nn
\eeq
\end{widetext}
where the local action $\bar{S}_0[a^*,a]$ is given by
\beq
&&\bar{S}_0[a^*,a]=\int d\t\sum_{i\in \mathcal{A}\oplus\mathcal{B} } a_i^*(\t)(\hb\partial_\t-\mu)a_i(\t)\nn\\
&&+\f{1}{2}U\sum_{i\in \mathcal{A}\oplus\mathcal{B}} a_i^*(\t)a_i^*(\t)a_i(\t) a_i(\t).  \nn
\eeq
We then make use of the cumulant expansion formula
\beq
\ex{e^{A_i}}=e^{\ex{A_i}+\f{1}{2}(\ex{A_i^2}-\ex{A_i}^2)+\ldots}\nn
\eeq
to expand the partition function $Z=\int \mathcal{D}\psi^* \mathcal{D}\psi\exp\{(-1/\hb) S_{eff}[\psi^*,\psi]\}$ in powers of $(\psi_i(\t),\psi_i^*(\t))$ to obtain the effective action $S_{eff}[\psi^*,\psi]$. Here, the expectation value of the field $\ex{A_i}_{\bar{S}_0}$ taken with respect to the weight $\exp\{(-1/\hb)\bar{S}_0[a^*,a]\}$ is defined in the usual way:
\beq
\ex{A_i  }_{\bar{S}_0}\equiv \int \mathcal{D}a^* \mathcal{D}a \,A_i\,  \exp \biggl\{ -\f{1}{\hb}\bar{S}_0[a^*,a]\biggr\}.\nn
\eeq
Close to the phase transition, where the Mott field vanishes, we keep only terms up to quadratic order in the cumulant expansion to get
\beq
&&S_{eff}[\psi^*,\psi]\approx \int d\t\sum_{<i,j>}\biggl[\psi_i^*(\t)\chi_{ij}\psi_j(\t)\nn\\
&&-\f{1}{2\hb} \biggl\langle \biggl(\psi^*_i(\t) \chi_{ij} a_j(\t)+ a_i^*(\t) \chi_{ij} \psi_j(\t) \biggr)^2  \biggr\rangle_{\bar{S}_0}  \biggr]. \nn
\eeq
We note that the expectation values with odd numbers of fields $(a_i^*,a_i)$ vanish. Furthermore, the local nature of the action $\bar{S}_0$ results in the identities
\beq
&&\ex{a_i(\t) a_j^*(\t')}_{\bar{S}_0}=\d_{ij}\ex{a(\t) a^*(\t')}_{\bar{S}_0},\nn\\ &&\ex{a_i(\t) a_j(\t')}_{\bar{S}_0}=\ex{a^*_i(\t)a^*_j(\t')}_{\bar{S}_0}=0.\nn
\eeq
By going to the momentum space, where the Mott fields are expressed as $\psi_{i\in \mathcal{A}}(\t)=\sum_{\tb{k}} a_\tb{k} (\t)\exp(i\tb{k} \cdot \tb{r}_i)$ and $\psi_{i\in \mathcal{B}}(\t)=\sum_\tb{k} b_\tb{k} (\t)\exp[i\tb{k}\cdot (\tb{r}_i+\tb{e}_1)])$, and simplifying, the effective action becomes
\beq
&&S_{eff}[\psi^*,\psi]=-\int d\t\sum_{\tb{k}}\biggl[\e_\tb{k} a^*_{\tb{k}}(\t) b_{\tb{k}}(\t)+\e_\tb{k}^* b^*_{\tb{k}}(\t) a_{\tb{k}}(\t)\biggr]\nn\\&&\tr{\ \ \ \ \ \ \ \ \ \ \ \ \ \ \ \ \ \ }-\f{1}{\hb}\sum_\tb{k}\int d\t d\t'\ex{ a(\t)a^*(\t') }_{\bar{S}_0}\nn\\&&\tr{\ \ \ \ \ \ \ \ \ \ \ \ \ \ \ \ \ \ } \times\biggl[a_\tb{k}^*(\t)a_\tb{k}(\t')+b_\tb{k}^*(\t)b_\tb{k}(\t')\biggr]|\e_\tb{k}|^2.\nn
\eeq
Now, since the Mott state with vanishing hopping is spanned by Fock states with fixed number of particles, the two-points Green function $\ex{a(\t)a^*(\t')}$ can be evaluated exactly to yield
\beq
\ex{a(\t)a^*(\t') }_{\bar{S}_0}&=&\theta(\t-\t')(n+1)e^{-(-\mu+nU)(\t-\t')/\hb}
\nn\\&&+\theta(\t'-\t)ne^{(-\mu+(n-1)U)(\t'-\t)/\hb}.\nn
\eeq
Substituting this expression into the effective action, expanding the fields in the Matsubara frequencies
\beq
&&a_\tb{k}(\t)=\f{1}{\sqrt{\hb \B}}\sum_{m} e^{-i\o_m\t}a_{\tb{k},\o_m},\nn\\ &&b_\tb{k}(\t)=\f{1}{\sqrt{\hb \B}}\sum_{m} e^{-i\o_m\t}b_{\tb{k},\o_m},
\eeq
and using a representation for the step function
\beq
\th (\t-\t')=-\int_{-\infty}^{\infty}\f{d \varsigma}{2\pi i}\f{e^{-i\varsigma(\t-\t')}}{\varsigma+i\eta},\nn
\eeq
we finally obtain the effective action up to quadratic order
\begin{widetext}
\beq
S_{eff}[a_{\tb{k},\o_m}^*,a_{\tb{k},\o_m},b_{\tb{k},\o_m}^*,b_{\tb{k},\o_m}]&=&-\sum_{\tb{k},m}\biggl[\e_\tb{k} a^*_{\tb{k},\o_m} b_{\tb{k},\o_m}+\e_\tb{k}^* b^*_{\tb{k},\o_m} a_{\tb{k},\o_m}+|\e_\tb{k}|^2 f_{\o_m}( a_{\tb{k},\o_m}^*a_{\tb{k},\o_m}+b_{\tb{k},\o_m}^*b_{\tb{k},\o_m})\biggr]\nn\\
&\equiv&\sum_{\tb{k},m}
\bem
a_{\tb{k},\o_m}\\b_{\tb{k},\o_m}
\eem^\dag
\biggl(-\hb\tb{G}^{-1}(\tb{k},i\o_m)
\biggr)\bem
a_{\tb{k},\o_m}\\b_{\tb{k},\o_m}
\eem\nn
\eeq
\end{widetext}
where
\beq
f_{\o_m}=\f{n+1}{-i\hb\o_m-\mu+n U }+\f{n}{i\hb\o_m+\mu-(n-1)U}.\nn
\eeq
In order to determine the excitation spectrum, we perform an analytic continuation in the frequency space $i\hb\o_m\rightarrow \hb\o$ and locate the poles of the Green function $\tb{G}(\tb{k},i\o_m)$. In this case, it amounts to solving $\tr{det} [\tb{G}^{-1}]=0$, or
\beq
|\e_\tb{k}|^2(|\e_\tb{k}|^2 f_{\o}^2-1)=0.
\eeq
We then obtain two branches of the quasiparticle and quasihole spectra in the Mott state,
\beq
\hb\o_1^{qp,qh}&=&\f{1}{2}\biggl(-|\e_\tb{k}|-2 \mu+(2n-1)U\nn\\&&\pm\sqrt{|\e_\tb{k}|^2-(4n+2) |\e_\tb{k}| U+U^2}\biggr),\nn
\eeq
\beq
\hb\o_2^{qp,ph}&=&\f{1}{2}\biggl(|\e_\tb{k}|-2 \mu+(2n-1)U\nn\\&&\pm\sqrt{|\e_\tb{k}|^2+(4n+2) |\e_\tb{k}| U+U^2}\biggr).\nn
\eeq
Since the quasiparticle and quasihole are produced pairwise in the Mott state, we look for the difference in the quasiparticle-quasihole spectra to obtain the excitation spectrum
\beq
E_{\tb{k}}=\sqrt{|\e_\tb{k}|^2-(4n+2) |\e_\tb{k}| U+U^2},
\eeq
where the single-particle spectrum $|\e_\tb{k}|$ depends implicitly on the staggered-flux strength. At a fixed filling, the SF-MI transition is then located at the point where the gap vanishes. Hence, by fixing $n=1$ and evaluating $E_{\tb{k}}=0$ we find the boundaries between the SF and the MI phases [see Fig.~\ref{Fig.4}(b)]. Thus, the SF-MI transition has been generalized to the case where the critical coupling $(U/4J)_c$ also depends on the strength of the staggered flux.

\section{Experimental Signatures of the distinct superfluids} \label{sec6}
\begin{figure}
\includegraphics[scale=.28, angle=0, origin=c]{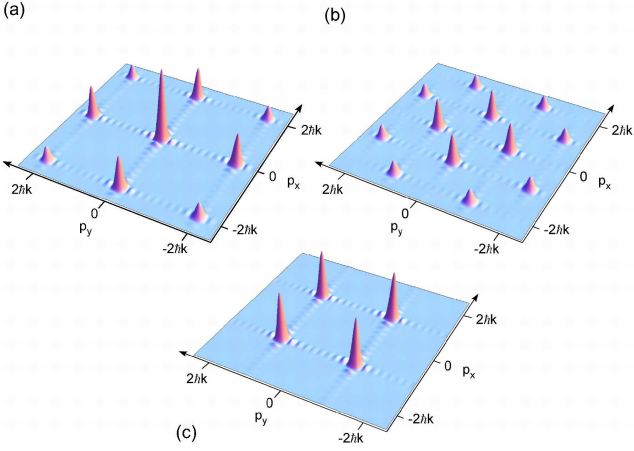}
\caption{\label{Fig.5}(Color online) Momentum spectra for (a) the uniform $(1,1,1,1)$ superfluid, (b) the staggered--vortex $(1,\pm i,-1,\mp i)$ superfluid, (c) and the staggered--sign (1,-1,1,-1) superfluid.}
\end{figure}
A simple method to distinguish the bosonic superfluids experimentally is to image their momentum distributions. This is achieved by allowing the system to expand ballistically after turning off the confining potential and subsequently imaging the atomic density with standard techniques. For sufficiently long expansion times, the atomic density reflects the initial momentum distribution. The momentum distribution of the condensed atoms is given by the quantity
\beq
\ex{\Psi^\dag(\tb{k})\Psi(\tb{k})}&=&|w(\tb{k})|^2\biggl|\sum_{\tb{R}\in\square}e^{i\tb{k}\cdot\tb{R}}\biggr|^2 v(\tb{k})\, .
\eeq
The first factor accounts for the Fourier transform of the Wannier function $w(\tb{k})$. The second factor is the structure factor of the Bravais lattice spanned by the vectors $2\tb{e}_1, 2\tb{e}_2$; that is, the sum extends over all  \textit{shaded} plaquettes according to Fig.~\ref{Fig.1}(d). The third factor, $v(\tb{k})$, is the form factor of the elementary plaquette defined by
\beq
v(\tb{k})=\sum_{\nu,\mu=1}^4
e^{i\tb{k}\cdot(\vec{\iota}_\nu-\vec{\iota}_\mu)}\ex{a^\dag_\nu a_\mu}
\eeq
with $\vec{\iota}_\nu$ ($\nu=1,2,3,4$) indicating the positions of the four sites in the plaquette. The expectation values $\ex{a^\dag_\nu a_\mu}$ can be evaluated for either of the wave functions in Eqs.~(\ref{Har1}) and (\ref{Har2}). This task is considerably simplified by observing that in the limit of large lattices, these wave functions (after some algebra) can be expressed as products of coherent states formed at each lattice site,
\beq
\label{coherent}
\dr{\Psi} = \prod_{\tb{r}\in \square}  \prod_{\nu =1,2,3,4}   \dr{\sqrt{\bar{n}}\,\psi_\nu}_ {\nu,\tb{r}}\, ,
\eeq
where $\psi_\nu$ denotes the respective order parameter from Eq.~(\ref{order1}) or Eq.~(\ref{order2}) and $\dr{\sqrt{\bar{n}}\,e^{i \gamma}}_ {\nu,\tb{r}}$ denotes a coherent state at corner $\nu$ of plaquette $\tb{r}$ with an average $\bar{n}$ atoms and a phase $\gamma$. With the help of Eq.~(\ref{coherent}), the plaquette form factors $v_{0,\pm}(\tb{k})$ for the homogeneous $(1,1,1,1)$ and the staggered--sign $(1,-1,1,-1)$ superfluids,  and $v_{\pi,\pm}(\tb{k})$ for the staggered--vortex superfluids $(1,\pm i,-1, \mp i)$ are evaluated to give
\beq
v_{0,+}(\tb{k})&=&4 \cos^2\biggl(k_x\f{\l}{4}\biggr) \cos^2\biggl(k_y\f{\l}{4}\biggr), \nn \\
v_{0,-}(\tb{k})&=&4 \sin^2\biggl(k_x\f{\l}{4}\biggr) \sin^2\biggl(k_y\f{\l}{4}\biggr), \nn \\
v_{\pi,\pm}(\tb{k})&=&4\biggl[\sin^2\biggl((k_x+k_y)\f{\l}{4}\biggr)\nn\\&&
+\sin^2\biggl((k_x-k_y)\f{\l}{4}\biggr) \biggr]^2,
\eeq
with $k_x \equiv \tb{k} \cdot \,\hat x, k_y \equiv \tb{k}\cdot\, \hat y$. The resulting momentum spectra are shown in Fig.~\ref{Fig.5}. One recognizes the absence of the zero momentum peak for the staggered--vortex and the staggered--sign superfluids in Figs. 5(b) and 5(c). Whereas for the uniform phase lattice momentum and momentum are equal, the staggered-sign phase is composed of momentum components which differ from $\tb{k}=0$ by a primitive vector of the reciprocal lattice. The clearly different patterns of Bragg peaks permit a direct identification of the respective superfluid in experiments.

\section{Discussions and Conclusions}
\label{conc}
In this paper, a tight-binding model was studied provided by an optical square lattice subjected to a time-dependent modulation, which excites staggered currents. Two different methods were used to show that the time-independent effective description of the system is equivalent to the Hubbard model in the presence of a staggered magnetic field. Due to the sublattice degrees of freedom, the single-particle spectrum of the model presents several interesting features, such as two inequivalent conical points in the energy band and distinct energy minima that depend on the magnitude of the staggered magnetic field. Then two cases were considered, first an optical lattice loaded with spinless fermions and then a lattice loaded with bosons.

When the optical lattice is half-filled with spinless fermions, the low energy excitations are governed by a Dirac-like dispersion. The problem is then reminiscent of graphene. However, here the cones are in general anisotropic, with the Fermi velocity controlled by the staggered flux. This feature cannot be easily implemented in graphene, where the Dirac cones arise due to the hexagonal lattice geometry. Nevertheless, anisotropic Dirac cones can be obtained by growing the graphene layer on top of a periodically patterned potential. This method, however, imprints a fixed anisotropy which cannot be tuned at will as in the case of the fermionic cold-atom system.

When the optical lattice is loaded with bosons, novel superfluid states arise because the location of the minimum of the single-particle spectrum depends on the staggered magnetic field: for a staggered flux $-\pi<\phi < \pi$, the minimum lies at ${\bf k}_0 = 0$ and for weak interactions a conventional uniform superfluid phase is realized. For $\pi < \phi < 3\pi$, the minimum lies at ${\bf k}_0 = (\pm \pi/d, \pm \pi/d)$ and thus a finite-momentum superfluid is realized. This phase corresponds to a vortex-antivortex square lattice. An analog phase, however, with opposite rotational direction is realized for $-3\pi<\phi<-\pi$. Finally, for $-4\pi<\phi< -3\pi$ and $3\pi<\phi<4\pi$, a staggered-sign superfluid phase emerges with an order parameter that has opposite signs for the two sublattices. We note that the time-modulation technique used here to generate the staggered magnetic field does not permit to access flux values $\phi$ outside the range $[-2\pi,2\pi]$, where the conventional tunnelling strength $J_0$ is positive. It was then shown that the different superfluid states are separated from each other by first-order phase boundaries within the mean-field analysis. For larger $U$, a second-order phase transition to a Mott-insulator arises, where the staggered flux renormalizes the critical coupling. Finally, the distinct experimental signatures of the superfluids that could be observed in standard ballistic expansion experiments were discussed.

\section*{Acknowledgements} This work was partially supported by the Netherlands Organization for Scientific Research (NWO).
A. H. acknowledges support by DFG (He2334/10-1) and Landesexzellenzcluster ``Frontiers in Quantum Photon Science". We are grateful to H. T. C. Stoof and B. Dou\c{c}ot  for fruitful discussions. We would also like to thank O. Tieleman for a careful reading of the manuscript.

\end{document}